\documentclass[12pt,preprint]{aastex}

\begin{document}

\title{The Deep and Low-Mass-Ratio Contact Binary CSS\,J022914.4$+$044340 with A Luminous Additional Companion}

\author{Liang Liu\altaffilmark{1,2,3,4}, and Xu-Zhi Li\altaffilmark{1,2,3,4}}\singlespace

\altaffiltext{1}{Yunnan Observatories, Chinese Academy of Sciences, 396 Yangfangwang, Guandu District, Kunming, 650216, P. R. China (e-mail: LiuL@ynao.ac.cn)}
\altaffiltext{2}{Key Laboratory for the Structure and Evolution of Celestial Objects, Chinese Academy of Sciences, 396 Yangfangwang, Guandu District, Kunming, 650216, P. R. China}
\altaffiltext{3}{Center for Astronomical Mega-Science, Chinese Academy of Sciences, 20A Datun Road, Chaoyang District, Beijing, 100012, P. R. China}
\altaffiltext{4}{University of Chinese Academy of Sciences, Yuquan Road 19\#, Sijingshang Block, 100049 Beijing, China}

\begin{abstract}
The first $B$-, $V$-, $R_c$-, and $I_c$-band light curves of CSS\,J022914.4$+$044340 are presented and analyzed. It is found that CSS\,J022914.4$+$044340 is a low mass ratio ($0.198\pm0.005$) deep ($63.7\pm7.9$\,\%) contact binary, indicating that it has been already at the end evolutionary stage of tidally-locked evolution via magnetized wind. Because of the totally eclipsing character, the photometric solutions are reliable. The temperature and the metallicity are determined from the spectroscopic data as $T=5855\pm15$\,K, and [Fe/H] $=-0.842\pm0.031$, respectively. Based on the parallax of $Gaia$ EDR3, the physical parameters of CSS\,J022914.4$+$044340 are estimated as $M_1=1.44^{+0.25}_ {-0.22}\,\rm{M_{\odot}}$, $M_2=0.29^{+0.05}_ {-0.05}\,\rm{M_{\odot}}$, $R_1=1.26^{+0.08}_ {-0.06}\,\rm{R_{\odot}}$, $R_2=0.65^{+0.03}_ {-0.04}\,\rm{R_{\odot}}$, $L_1=1.718^{+0.186}_ {-0.191}\,\rm{L_{\odot}}$, $L_2=0.416^{+0.039}_ {-0.050}\,\rm{L_{\odot}}$. Combined the fraction in light of the third body via the photometric solution (54\%), the luminosity of the third body is estimated as $2.705\,\rm{L_{\odot}}$. The third body may be inferred as a subgiant. Thus, it is explained that why the primary component of CSS\,J022914.4$+$044340 has higher mass among the similar systems, and why its metallicity is so poor.
\end{abstract}

\keywords{Binaries : eclipsing -- Binaries : close -- Stars: individuals (CSS\,J022914.4$+$044340) -- Stars: evolution}
\section{Introduction}
A contact binary is a close binary system whose components overflow their respective Roche lobes \citep{Kopal1959} and share a common envelope (CE). The common envelope is radiative if the system is composed of O, B, or maybe A spectral type stars, or else it is convective if the system is composed of the later spectral type stars. \cite{Lucy1968a,Lucy1968b} had put forward a wide accepted convective common envelope (CCE) model for the W UMa-type contact binaries which contain two unevolved low mass components. Because of the tight orbit, contact binaries have the shortest periods, the lowest angular momenta and the strongest interactions among close binaries. They are very important for studying the processes of the free mass transfer and the energy transfer, the interaction of components, and characters of the common envelope (CE). Due to the rapid rotating later-type component, contact binaries usually showed magnetic activities. In these cases, the light curves would be distorted by dark spots so-call the O'Connell effect  \citep{O'Connell1951}. Some of them, the dark spots were found changing within a short timescale (e.g., GN Boo, \citealt{Wangetal2015}; V789 Her, \citealt{Lietal2018}). More important, contact binaries were associated with some special celestial systems. The progenitor of the luminous red nova (LRN) V1309 Sco \citep[e.g.,][]{Tylendaetal2011,Stepien2011} would be a deep and low-mass-ratio contact binary (DLMRCB) \cite{Zhuetal2016}. The Am-eclipsing binary V2787 Ori is also a shallow contact and extreme mass ratio contact binary \citep{Tianetal2019}. V53, a member of the globular cluster M4, could be a blue straggler (BS) \citep{Lietal2017}. And recently, \cite{Ferreiraetal2019} reported that the post-burst V1309 Sco, a merger from a contact binary, is located in the BS region.

There is a period cutoff of contact binaries. This phenomenon was discovered by \cite{Rucinski1992}. According to the data at that time, he found the value of the period cutoff is about 0.22\,days and he suggested that the limitation could be due to the fully convective structure of the M-type component. Later, \cite{Stepien2006} explained this period limit as the shorter period detached binaries have not evolved into contact binaries within the age of the Universe, yet. \cite{Lietal2019} collected 55 such samples and suggested that in some cases, the effect of the third body should be taken into account except for those two reasons. They also suggested statistical value for the period cutoff as 0.1763835\,days. And very recently, \cite{ZhangQian2020} suggested this value as 0.15\,days. More information about the short period cutoff of contact binaries could be found in the review \citep[][and the references therein]{Qianetal2020}.

Contact binaries are usually formed from detached binaries \cite[e.g.,][]{Eggleton2012} via angular momentum loss (AML) or via the third bodies through the Kozai mechanism \citep{Kozai1962}. According to the spectra provided by LAMOST, \cite{Qianetal2017,Qianetal2018} found that the EA-type binaries have higher metallicities than the EW-type binaries and they suggest that the long-period EWs should be formed by the EAs. During the contact evolutionary phase, there are at least two very important evolutionary stages for contact binaries, namely the marginal contact phase and the deep and low-mass-ratio (DLMR) contact phase \citep{Qianetal2020}. The marginal contact phase was predicted by the thermal relaxation oscillation (TRO) theory \citep{Lucy1976,Flannery1976,RobertsonEggleton1977}. This theory assumes that matter can be transferred via the inner Lagrangian point while energy can be exchanged through the CCE. Mass transfer changes the orbit and energy transfer varies the surface temperature. Such processes make contact binaries oscillate between the semi-detached phase and the shallow contact phase. If the marginal contact binary has an appropriate rate of angular momentum loss (AML) so that it counteracts the increased orbit causing by mass transfer, the system could keep in the zero-age and shallow contact phase \citep[e.g.,][]{Rahunen1981}. Based on investigations of the period variations, it is suggested that contact systems could oscillate around a critical mass ratio on the thermal timescales \citep[e.g. $q_{\rm{crit}}=0.4$,][]{Qian2001,Qian2003}. Had undergone many times of such cycles, if the rate of AML increased, contact binaries would evolve into the DLMR contact phase. A contact binary that has mass ratio $q \leqslant 0.25$ and fill-out factor $f\geqslant 50\%$ is suggested to called a DLMRCB \cite[e.g.,][]{Qianetal2005,Qianetal2006,YangQian2015}. DLMRCBs are believed to merge into fast rotating single stars as FK Com-type stars or BSs because of dynamical instability causing by the extreme mass ratio or by the great depth of contact \citep[e.g.,][]{EggletonKiseleva-Eggleton2001,Eggleton2012}.

CSS\,J022914.4$+$044340 (CRTS\,J022914.4$+$044340) was found as an EW-type eclipsing binary by \cite{Drakeetal2014}. It was taken as a candidate of DLMRCB because of the flat bottom profile on its minimum and low amplitude of its light variations. This target was observed by LAMOST for serval times. According to latest data release of LAMOST \citep{Luoetal2019}, we have T $=5767.91\pm85.13$\,K and [Fe/H]$=-0.922\pm0.082$. We can also find this target in the $Gaia$ DR2 and $Gaia$ EDR3. Then we have T $=5864.25^{+260.41}_ {-79.25}$\,K, G $=14.2893\pm0.0072$\,mag \citep{GaiaCollaboration2018}, and Parallax $=0.6250\pm0.0210$\,mas \citep{GaiaCollaboration2020}. This Parallax corresponds to a distance of $1600.0^{+55.6}_ {-52.0}$\,pc. This paper is organized as follows. In section 2, the photometric and spectroscopic observations for this target are introduced briefly. In section 3, the light curves solution is obtained via the W-D code. In section 4, the physical parameters are estimated. And the last section is about discussions and conclusion.

\section{Photometric and spectroscopic observations and data reduction}
The $B$-, $V$-, $R_c$-, and $I_c$-band light curves of CSS\,J022914.4$+$044340 were carried out in two nights on October 26 and 28, 2017, with an Andor DZ936 2K CCD camera attached to the 85\,cm reflecting telescope at the Xinglong Station of the National Astronomical Observatories of the Chinese Academy of Sciences. The observational system has a standard Johnson-Cousins-Bessel multi-color CCD photometric system built on the primary focus \citep{Zhouetal2009}, generating an effective field of view equal to $33 \times 33$\,arcmin$^2$. The integration times were 80\,s for the $B$-band, 50\,s for the $V$-band, 40\,s for the $R_c$-band, and 30\,s for the $I_c$-band, respectively. During the observation, the weather was clear and the seeing was about $1.2^{\prime\prime}$.

To avoid the cumbersomely standard photometry, the differential photometry method was used to measure the light curves of the variable star. The effect of the atmospheric extinction can be easily eliminated with this method by selecting suitable comparison and check stars. In this reduction, 2MASS\,J02285932$+$0444322 and 2MASS\,J02290506$+$0441298 were chosen as comparison star and check star, respectively. According to the $Gaia$ DR2 \citep{GaiaCollaboration2018}, we have T $=5431.93$\,K, G $=13.7433$\,mag for the comparison, and T $=5375.50$\,K, G $=14.7731$\,mag for the check star, respectively. The observed data were reduced by the IRAF, with the bias and flat corrections. PHOT was used to obtain the instrument magnitudes of the chosen stars. All frames are measured. The measuring errors of the magnitude ($\sim0.02$\,mag) mainly come from the lower S/N because of the faint target. The phases of CSS\,J022914.4$+$044340 are calculated according to the ephemeris that Min.I~$=2458055.17640+0^d.306139{\times}E$. $T_0$ of 2458055.17640 fitted by a parabola with the Least Square Method is the primary minimum of this observation, and $P_0$ of $0^d.306139$ is from the study of \cite{Qianetal2017}. One can find the corresponding light curves in Figure~\ref{fig:light curves_figure}. Several new times of minima determined via the parabolic fitting method are listed in Table~\ref{tab: New minima}.

\begin{figure}
	\centering
	\includegraphics[angle=0,scale=.55]{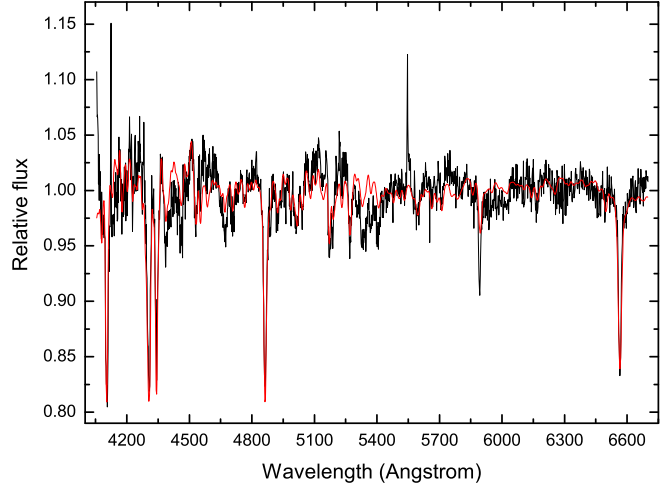}
\caption{Spectrum of CSS\,J022914.4$+$044340. Black solid line is the observed spectrum, while the red solid line is the fitting. The spectrum was observed by the 2.4-m telescope in Lijiang Gaomeigu Station, with the Grism-14 and the $2''.5$ long-slit. The exposure time was 3600\,s.}
    \label{fig:spec}
\end{figure}

We determine the temperature of CSS\,J022914.4$+$044340 by using of the Yunnan Faint Object Spectrograph and Camera (YFOSC). The spectroscopic observations were carried out on January 17, 2020, with the 2.4-m telescope in Lijiang Gaomeigu Station. The filter is Grism-14 covering the wavelength from 320 to 750\,nm, while the long-slit is about $2''.5$. More details of this spectroscopic system can be found in the paper of \cite{Fanetal2015}. Because our target is faint, an exposure time of 3600\,s was adopted. We employed IRAF to eliminate the affects of bias and flat on the picture. Then, we used the APEXTRACT package to extract the spectra. The comparison spectra for wavelength calibration was from the He-Ne lamp. Finally, we fit the reduced spectrum with the package of University of Lyon Spectroscopic analysis Software (ULySS) \citep{Kolevaetal2009,Wuetal2011}. UlySS is a full-spectrum fitting method for the analysis of stellar atmospheric parameters and stellar population spectra. In this method, chi-square maps are used to study the degeneracies; convergence maps are used to determine the local minima; and Monte Carlo simulations are used to estimate the errors. The observed spectrum of CSS\,J022914.4$+$044340 and the corresponding fitting are list in Figure~\ref{fig:spec}. According to this fitting, we obtain that $T_{\rm{eff}}=5855\pm15$\,K and [Fe/H] $=-0.842\pm0.031$.

\section{Light curves solution}
\begin{figure}
	\centering
	\includegraphics[angle=0,scale=.8]{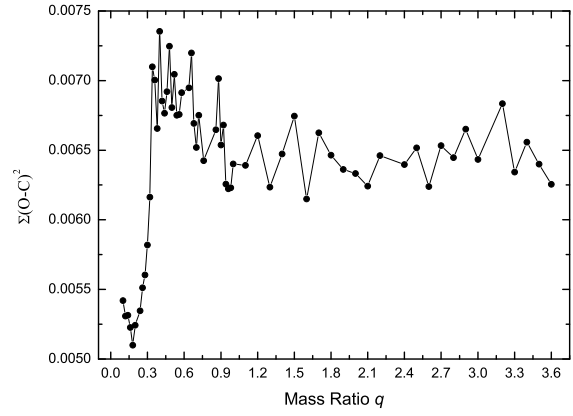}
\caption{The relation between $q$ and the residual sum of squares for CSS\,J022914.4$+$044340. The optimally initial $q$ is about 0.18.}
    \label{fig:qs}
\end{figure}

To determine the photometric elements and further to understand the geometrical structure and the evolutionary state of CSS\,J022914.4$+$044340, the latest version of the W-D code \citep{WilsonDevinney1971, Wilson1979, Wilson1990, Wilson2008, Wilson2012, vanHammeWilson2007, Wilsonetal2010, WilsonvanHamme2014} are employed to analyze the multi-color light curves. The $q$-search method is adopted to find an initial value of $q$. The result of the $q$-search which is shown in Figure~\ref{fig:qs} suggests an initial value of $q$ as 0.18.

\begin{figure}
	\centering
	\includegraphics[angle=0,scale=.7]{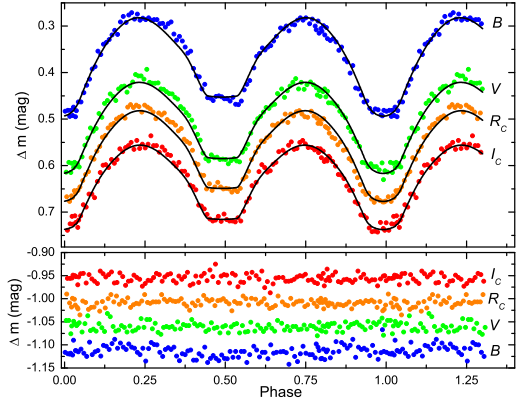}
 \caption{In the upper panel, the color solid circles refer to the observed differential light curves for CSS\,J022914.4$+$044340, while the black solid lines refer to the theoretical light curves. Different colors denote different filters. The $I_c$-band light curve is shifted with $+0.05$\,mag. The light curves in the lower panel refer to the comparison star minus the check star (C-Ch). The symbols are the same as those in the upper panel.}
    \label{fig:light curves_figure}
\end{figure}

That the temperature of CSS\,J022914.4$+$044340 is about $T_{\rm{eff}}=5855\pm15$\,K suggests a convective equilibrium for this system, corresponding a bolometric albedo of $A_1=A_2=0.5$ \citep{Rucinski1969} and a gravity-darkening coefficient of $g_1=g_2=0.32$ \citep{Lucy1967}. We adopt the square root law to handle the effect of the limb-darkening, because the given errors for this law on that temperature are the smallest. The values of the limb-darkening coefficients come from \cite{ClaretBloemen2011}. The adjusted parameters are the mass ratio $q$, the mean temperature of star\,2, $T_2$, the monochromatic luminosity of star\,1 and the dimensionless potential of star\,1 (mode\,3, $\Omega_1=\Omega_2$). The related limb-darkening coefficients varies with $T_2$. We also adjust $l_3$ as the luminosity of third body in the sight line, finding that the third light is more than half the luminosity of the system. The solved photometric parameters are listed in Table~\ref{tab:solution_table fitted}, while the fitted light curves are shown in Figure~\ref{fig:light curves_figure}. It is illustrated with Figure~\ref{fig:light curves_figure} that CSS\,J022914.4$+$044340 is a totally eclipsing binary.

\section{Physical parameters estimation}
We have introduced how to estimate the absolute parameters ($M, R, L$) with errors via the parallax given by $Gaia$ in a very recent paper \citep{Liuetal2020}. We know that the Parallax of CSS\,J022914.4$+$044340 is $0.6250\pm0.0210$\,mas \citep[$Gaia$ EDR3,][]{GaiaCollaboration2020}, being corresponding to a distance of $1600.0^{+55.6}_ {-52.0}$\,pc, yielding a modulus of $11.021^{+0.074}_ {-0.072}$\,mag. We also know the V-band magnitude as $13.835\pm0.051$\,mag \citep[URAT1 Catalog,][]{Zachariasetal2015} for our comparison star (2MASS\,J02285932$+$0444322). From the light curves (the upper panel of Figure~\ref{fig:light curves_figure}), we determine that the maximum V-band magnitude of CSS\,J022914.4$+$044340 is about 14.256\,mag while the corresponding minimum magnitude is about 14.456\,mag. Thus we infer the absolute magnitude as $3.235^{+0.074}_ {-0.072}$\,mag in V-band at the phase of 0.25 or 0.75. According to \cite{WortheyLee2011}, we have a bolometric correction coefficient for V-band as $-0.121\pm0.050$\,mag under the conditions of $T=5855\pm15$\,K and [Fe/H] $=-0.842\pm0.031$. Therefore, we estimate the absolute bolometric magnitude of CSS\,J022914.4$+$044340 as $3.114^{+0.124}_ {-0.122}$\,mag. By using the formulae introduced in our paper earlier in this year \citep{Liuetal2020}, we have $M_1=1.44^{+0.25}_ {-0.22}\,\rm{M_{\odot}}$, $M_2=0.29^{+0.05}_ {-0.05}\,\rm{M_{\odot}}$, $R_1=1.26^{+0.08}_ {-0.06}\,\rm{R_{\odot}}$, $R_2=0.65^{+0.03}_ {-0.04}\,\rm{R_{\odot}}$, $L_1=1.718^{+0.186}_ {-0.191}\,\rm{L_{\odot}}$, $L_2=0.416^{+0.039}_ {-0.050}\,\rm{L_{\odot}}$ for each component of CSS\,J022914.4$+$044340. These errors are derived from the error transfer formula. All estimated parameters are listed in Table~\ref{tab:physical parameters_table}.

\section{Discussion and Conclusion}
\begin{figure}
	\centering
	\includegraphics[angle=0,scale=1]{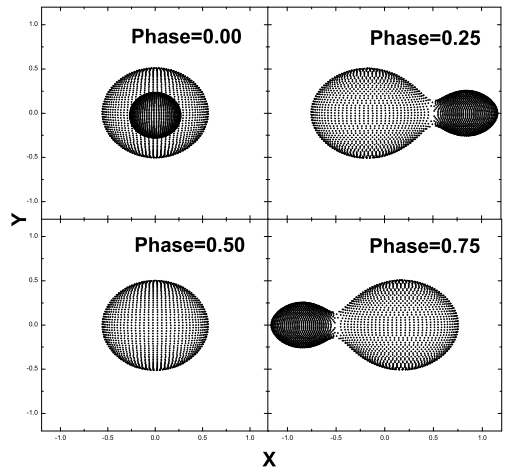}
\caption{The geometric structure of CSS\,J022914.4$+$044340 in different phases. It is obviously a totally eclipsing contact binary system.}
    \label{fig:structure_figure}
\end{figure}

Based on the solution of the multiple color light curves and the analysis of its uncertainties, we found that CSS\,J022914.4$+$044340 is an extreme mass ratio ($0.198\pm0.005$), deep contact binary ($63.7\pm7.9$\%). The 38-minute-last flat bottom in light curve, as well as the high inclination ($88^{\circ}.4\pm1^{\circ}.2$), indicates a total eclipse of this binary system (as shown in Figure~\ref{fig:structure_figure}). The photometric solutions would be reliable if the system were totally eclipsing \citep{TerrellWilson2005}. Based on that, the physical parameters have been estimated by its photometric solutions with its distance determined by $Gaia$ EDR3. To avoid the potential error brought from the temperature \citep[e.g.,][]{Liuetal2020}, we have determined the temperature of CSS\,J022914.4$+$044340 through the spectroscopic data obtained by the YFOSC. The determined value is $5855\pm15$\,K which is consistent with the value of $5767.91\pm85.13$\,K given by LAMOST DR5 \citep{Luoetal2019} and of $5864.25^{+260.41}_ {-79.25}$\,K given by $Gaia$ DR2 \citep{GaiaCollaboration2018}. Our derived value of metallicity as [Fe/H] $=-0.842\pm0.031$ is also similar to the value of $-0.922\pm0.082$ given by LAMOST DR5 \citep{Luoetal2019}.

According to the statistics \citep{Qianetal2017}, the peak of the metallicity for EW-type binaries is about $-0.3$. CSS\,J022914.4$+$044340 is close to the critical poorer metallicity boundary. This may be caused by the luminous third body. Our solved third body contributes over 54\% light in the system. Although it is very common that there is a third body in contact binaries \citep[e.g.,][]{PribullaRucinski2006,D'Angeloetal2006,Rucinskietal2007}, the luminous contribution of the third bodies being over 20\% is few. About 23\% luminosity of V345 Gem comes from the third body \citep{Yangetal2009}. These fractions are 25\% for MQ UMa \citep{Zhouetal2015}, 28\% for II UMa \citep{Zhouetal2016}, and 20\% for TZ Boo \citep{Christopoulouetal2011}. Except for II UMa, the orbital periods of these targets showed cyclic variations which would be caused by the light-travel-time effect (LTTE) \citep[e.g.,][]{LiaoQian2010}. According to \cite{D'Angeloetal2006}, II UMa contains a 0''.87-separation third body. Lack of the LTTE may be due to the long orbital period of the third body or it is a foreground/background star. The luminosity of the third body in CSS\,J022914.4$+$044340 is about 54\%. This fraction is very high. According to this fraction, the luminosity of this third body is estimated as $2.705\,\rm{L_{\odot}}$, corresponding to a main-sequence star with a mass of $1.3\,\rm{M_{\odot}}$ or a G5 subgiant with a mass of 1.1$\,\rm{M_{\odot}}$ \citep{Cox2000}. If it were former, its temperature as well as the metallicity would be a little higher than that of the Sun so that the corresponding values of CSS\,J022914.4$+$044340 should be lower than the derived valves mentioned above. The results would be reversed if it were later one. We also note that the mass of the primary is $M_1=1.44^{+0.25}_ {-0.22}\,\rm{M_{\odot}}$, but the temperature is $5855\pm15$\,K. The observed temperature of CSS\,J022914.4$+$044340 may be lower because of the existence of the conjectural subgiant third body. For the same reason, the observed metallicity may be lower, too. Hence, the real metallicity of CSS\,J022914.4$+$044340 may be not as low as the value of [Fe/H] $=-0.842\pm0.031$. Whether the parallax was affected by the third body is not clear.

On the other hand, contact binaries are taken as distance tracers because of the period-color correlation first discovered by \cite{Eggen1967}. The absolute-magnitude calibration of contact binaries was developed by \cite{Rucinski1994,Rucinski2000,Rucinski2004}, via the parallax data of the $Hipparcos$ Mission \citep{ESA1997} and the contact binaries in the open or globular clusters. By using the first $Gaia$ satellite Data Release \citep[DR1;][]{GaiaCollaboration2016a,GaiaCollaboration2016b}, \cite{MateoRucinski2017} improved this calibration. According to their formula
\begin{equation}
M_V=3.73-8.67({\rm{log}}P+0.40),
\end{equation}
substituting $P=0.306139$ days, we have $M_V=4.719$\,mag for CSS\,J022914.4$+$044340. However, the $M_V$ should be 4.103\,mag after excluding the affect of the third light. This difference may be caused by the reasons for bias of the temperature which was affected by the third light. DLMCBs with similar mass ratio to CSS\,J022914.4$+$044340 are collected in Table~\ref{tab:contact binaries with q around 0.2}. It is seen that CSS\,J022914.4$+$044340 is very similar to TZ Boo which has a spectroscopic mass ratio. However, the mass of the primary component of the former one is much larger. Again, this should be on account of the affect of the third light.

Based on the above analysis, we conclude that CSS\,J022914.4$+$044340 is a DLMRCB. It contains a perhaps subgiant third companion with a inferred mass of 1.1$\,\rm{M_{\odot}}$. The presence of the third body make the contact binary system seem cooler and metallicity poorer. The real mass of the primary component of CSS\,J022914.4$+$044340 should be a little less than 1.44$\,\rm{M_{\odot}}$.

\acknowledgments{We are grateful to the anonymous referee who has given very useful suggestions to improve the paper. We acknowledge the supports of the staff of the Xinglong 85-cm telescope and of the Lijiang 2.4-m telescope. We also acknowledge the supports from the Chinese Natural Science Foundation (No.\,11773066, and No.\,11933008), from the young academic and technology leaders project of Yunnan Province (No.\,2015HB098), and from the Open Project Program of the Key Laboratory of Optical Astronomy, National Astronomical Observatories, Chinese Academy of Sciences.}
\bibliographystyle{unsrt}
{}

\begin{table}
	\centering
\caption{New times of light minima for CSS\,J022914.4$+$044340 with the 85\,cm telescope at the Xinglong Station.}
\label{tab: New minima}

\begin{tabular}{llll}
\hline
J.D. (Hel. Day) & Error (d) & $Min.$ & Filter\\
\hline
2458053.18427  & 0.00092  & p   &  $B  $  \\
2458053.18284  & 0.00132  & p   &  $V  $  \\
2458053.18365  & 0.00064  & p   &  $R_C$  \\
2458053.18401  & 0.00115  & p   &  $I_C$  \\
2458055.17658  & 0.00048  & s   &  $B  $  \\
2458055.17673  & 0.00044  & s   &  $V  $  \\
2458055.17663  & 0.00044  & s   &  $R_C$  \\
2458055.17565  & 0.00044  & s   &  $I_C$  \\
\hline
\end{tabular}
\end{table}

\begin{table}
	\centering
\begin{tiny}
\caption{Photometric solutions for CSS\,J022914.4$+$044340.}
	\label{tab:solution_table fitted}
\begin{tabular}{lcl}
\hline
Parameters                      &  Photometric elements  &  errors      \\
\hline
$g_1=g_2$                       &     0.32               & assumed      \\
$A_1=A_2$                       &     0.50               & assumed      \\
$x_{\rm{1bol}},x_{\rm{2bol}}$   &     0.180,0.194        & assumed      \\
$y_{\rm{1bol}},y_{\rm{2bol}}$   &     0.537,0.523        & assumed      \\
$x_{1B},x_{2B}$                 &     0.491,0.532        & assumed      \\
$y_{1B},y_{2B}$                 &     0.395,0.353        & assumed      \\
$x_{1V},x_{2V}$                 &     0.229,0.257        & assumed      \\
$y_{1V},y_{2V}$                 &     0.606,0.581        & assumed      \\
$x_{1R_c},x_{2R_c}$             &     0.119,0.143        & assumed      \\
$y_{1R_c},y_{2R_c}$             &     0.650,0.632        & assumed      \\
$x_{1I_c},x_{2I_c}$             &     0.046,0.066        & assumed      \\
$y_{1I_c},y_{2I_c}$             &     0.638,0.623        & assumed      \\
Phase shift                     &     $-0.0092$          & $\pm0.0005$  \\
$T_1$~(K)                       &     5855               & $\pm15$      \\
$T_2$~(K)                       &     5747               & $\pm24$  	  \\
$q=M_2/M_1$                     &     0.198              & $\pm0.005$	  \\
$\Omega_{\rm{in}}$              &     2.2287             & --     	    \\
$\Omega_{\rm{out}}$             &     2.1021             & --     	    \\
$\Omega_1=\Omega_2$             &     2.1480             & $\pm0.0099$  \\
$i (^{\circ})$                  &     $88.4$             & $\pm1.2$ 	  \\
$L_1/(L_1+L_2)(B)$              &     0.8159             & $\pm0.0351$ 	\\
$L_1/(L_1+L_2)(V)$              &     0.8115             & $\pm0.0327$ 	\\
$L_1/(L_1+L_2)(R_c)$            &     0.8096             & $\pm0.0315$ 	\\
$L_1/(L_1+L_2)(I_c)$            &     0.8079             & $\pm0.0301$	\\
$l_3/(L_1+L_2+l_3)(B)$          &     0.5488             & $\pm0.0148$ 	\\
$l_3/(L_1+L_2+l_3)(V)$          &     0.5514             & $\pm0.0139$ 	\\
$l_3/(L_1+L_2+l_3)(R_c)$        &     0.5349             & $\pm0.0139$ 	\\
$l_3/(L_1+L_2+l_3)(I_c)$        &     0.5479             & $\pm0.0130$	\\
$r_1\rm{(pole)}$                &     0.5073             & $\pm0.0027$	\\
$r_1\rm{(side)}$                &     0.5595             & $\pm0.0042$	\\
$r_2\rm{(back)}$                &     0.5887             & $\pm0.0057$	\\
$r_2\rm{(pole)}$                &     0.2549             & $\pm0.0075$	\\
$r_2\rm{(side)}$                &     0.2687             & $\pm0.0095$  \\
$r_2\rm{(back)}$                &     0.3311             & $\pm0.0289$  \\
$f$ (\%)                        &    63.7                & $\pm7.9$     \\
\hline
\end{tabular}
\end{tiny}
\end{table}

\begin{table}
\caption{Estimated physical parameters of CSS\,J022914.4$+$044340, based on the resolution and the parallax from $Gaia$ EDR3.}
	\label{tab:physical parameters_table}
\begin{center}
\begin{tabular}{lcc}
\hline
Parameters                 &      Value           &        Range            \\
\hline
$T_1$ (K) 	               &       5855           &       5840 $\sim$ 5870         \\
$T_2$ (K) 	               &       5747           &       5723 $\sim$ 5771         \\
$M_1~\rm{(M_{\odot})}$     &       1.44           &       1.22 $\sim$ 1.69         \\
$M_2~\rm{(M_{\odot})}$     &       0.29           &       0.24 $\sim$ 0.34         \\
$R_1~\rm{(R_{\odot})}$     &       1.26           &       1.20 $\sim$ 1.34         \\
$R_2~\rm{(R_{\odot})}$     &       0.65           &       0.61 $\sim$ 0.68         \\
$L_1~\rm{(L_{\odot})}$     &       1.718          &      1.527 $\sim$ 1.904        \\
$L_2~\rm{(L_{\odot})}$     &       0.416          &      0.366 $\sim$ 0.455        \\
$A~\rm{(R_{\odot})}$       &       2.289          &      2.166 $\sim$ 2.418        \\
log$\,g_1$                 &       4.39           &       4.37 $\sim$ 4.42         \\
log$\,g_2$                 &       4.27           &       4.25 $\sim$ 4.29         \\
$M_{\rm{bol1}}$            &       4.22           &       4.10 $\sim$ 4.34         \\
$M_{\rm{bol2}}$            &       5.75           &       5.63 $\sim$ 5.87         \\
$m_V$                      &      14.255          &     14.204 $\sim$ 14.306      \\
Distance~(pc)              &       1600.0         &     1548.0 $\sim$ 1655.6       \\
$(m-M)_V$    		       &      11.021          &     10.949 $\sim$ 11.095       \\
$BC_V$                     &      $-0.121$        &   $-0.171$ $\sim$ $-0.071$     \\
$M_{\rm{bol}}$             &       3.982          &      3.863 $\sim$ 4.102        \\
\hline
\end{tabular}
\end{center}
Note: The value of $M_{\rm{bol}}$ dose not include the third light.
\end{table}

\begin{table}
\begin{tiny}
\caption{DLMRCBs with $q$ around 0.20.}
	\label{tab:contact binaries with q around 0.2}
\begin{tabular}{llc cccccc cc cc c}
\hline
Star                &  Period      &   $q$      &   $M_1$         &  $M_2$         &  $R_1$	        & $R_2$	         & $L_1$          &  $L_2$          & $f$     & i	         & $T_1$    & $T_2$  & Reference                            \\
                    &   (days)     & $M_2/M_1$  & (${M_{\odot}}$) &(${M_{\odot}}$) &(${R_{\odot}}$) &(${R_{\odot}}$) &(${L_{\odot}}$) &(${L_{\odot}}$)   & \%      &($^{\circ}$) &  (K)     &  (K)   &                                 \\
\hline
TZ Boo              &  0.2971599   & 0.207      & 0.99  	        & 0.21           &	1.08  	      & 0.56           & 1.260          &	0.330           & 52.5 	 & 85.5 	     & 5890	    & 5873   & (1)     \\
CSSJ022914.4+044340 &  0.306139    & 0.198      & 1.44  	        & 0.29           &	1.26  	      & 0.65           & 1.718          &	0.416           & 63.7 	 & 88.2 	     & 5855	    & 5744   & (2)                      \\
BO Ari              &  0.3182      & 0.209      & 1.14  	        & 0.24           &	1.16  	      & 0.61           & 1.470          &	0.440           & 50.3 	 & 85.7 	     & 5920	    & 6055   & (3)              \\
V1853 Ori           &  0.3830      & 0.203      & 1.23  	        & 0.25           &	1.38  	      & 0.71           & 2.490          &	0.690           & 50.0 	 & 83.2 	     & 6200	    & 6261   & (4)             \\
NSVS 6859986        &  0.38356914  & 0.208      & 1.87  	        & 0.39           &	1.63  	      & 0.84           & 1.620          &	0.430           & 86.4 	 & 89.0 	     & 5100	    & 5100   & (5)       \\
TYC 3836-0854-1     &  0.41556601  & 0.190      & 1.200 	        & 0.228          &	1.46  	      & 0.75           & 3.091          &	0.795           & 79.4 	 & 77.5 	     & 6332	    & 6292   & (6)            \\
MQ UMa              &  0.47606620  & 0.195      & 				    & 		           &                &                &                &                 & 82.0 	 & 65.6 	     & 6352	    & 6224   & (7)             \\
DN Aur              &  0.6169      & 0.205      & 1.44  	        & 0.30           &	1.98  	      & 1.01           & 7.570          &	1.880           & 53.9 	 & 76.9 	     & 6830	    & 6750   & (8)        \\
HV UMa              &  0.7108      & 0.190      & 2.84  	        & 0.54           &	2.62  	      & 1.18           & 17.220         & 2.950           & 61.9 	 & 57.3 	     & 7300	    & 7000   & (9)             \\
\hline
\end{tabular}
Reference: (1) \cite{Christopoulouetal2011}; (2) This paper; (3) \cite{YangQian2015}; (4) \cite{Samecetal2011}; (5) \cite{Kjurkchievaetal2019}; (6) \cite{Liaoetal2017}; (7) \cite{Zhouetal2015}; (8) \cite{Goderyaetal1996}; (9) \cite{Csaketal2000}.
\end{tiny}
\end{table}

\end{document}